\documentstyle[nato,psfig,namedreferences]{crckapb}

\begin{opening}

\title{RECONSTRUCTING THE STAR FORMATION HISTORY\protect\\
        OF THE GALAXY }

\author{X. HERNANDEZ}
\institute{Osservatorio Astrofisico di Arcetri,\\
           Largo E. Fermi 5, 50125 Firenze, Italy}
\author{D. VALLS--GABAUD}
\institute{Laboratoire d'Astrophysique, UMR CNRS 5572,\\
	   Observatoire Midi-Pyr\'en\'ees, \\
	   14 Av. E. Belin, 31400 Toulouse, France}
\author{G. GILMORE}
\institute{Institute of Astronomy,\\
	   Madingley Road, Cambridge CB3 0HA, UK}

\end{opening}

\runningtitle{GALACTIC STAR FORMATION HISTORY}

\begin{document}

\begin{abstract}

The evolution of the star formation rate in the Galaxy is one of the
key ingredients quantifying the formation and determining the chemical
and luminosity evolution of galaxies. Many complementary methods exist
to infer the star formation history of the components of the Galaxy,
from indirect methods for analysis of low-precision data, to new exact
analytic methods for analysis of sufficiently high quality data. We
summarise available general constraints on star formation histories,
showing that derived star formation rates are in general comparable to
those seen today.  We then show how colour-magnitude diagrams of
volume- and absolute magnitude-limited samples of the solar
neighbourhood observed by Hipparcos may be analysed, using variational
calculus techniques, to reconstruct the local star formation
history. The remarkable accuracy of the data coupled to our
maximum-likelihood variational method allows objective quantification
of the local star formation history with a time resolution of $\approx
50$Myr.  Over the past 3Gyr, the solar neighbourhood star formation
rate has varied by a factor of $\sim$4, with characteristic timescale
about 0.5Gyr, possibly triggered by interactions with spiral arms.

\end{abstract}

\section{Star Formation Rates: Qualitative Deductions}

Star formation rates and histories can be estimated in special cases
from a combination of chemical evolution models and the total stellar
mass formed into stars. Basically, this exploits the stellar
evolutionary, and Type~I supernova timescales for element production.
One requires that stars formed at a rate consistent with their
chemical distributions, whether a $\delta$-function, a range in the
products of both Type~I and Type~II supernaova, products of single
supernovae with time for efficient mixing of the ISM, or whatever.
Combining these albeit crude estimates of the {\sl duration} of star
formation with a calculation of the stellar mass formed, provides a
star formation rate. Perhaps surprisingly, given the crude
calculation, such derived rates are both similar to those determined
more accurately today, and are all quite low.

\begin{table}[h]
\begin{center}
\caption{A summary of star formation rates, and durations of star
formation, in some Galactic stellar populations. These values are
derived from combination of chemical element scatter and masses.}
\vskip 10pt
    \leavevmode    
    \footnotesize
    \begin{tabular}[tbh]{|l|c|c|}
\hline && \\[-4pt]
{ Population }  & Duration & Formation Rate \\
\hline && \\[-4pt]
  &  (years) & ${\mathcal M}_{\odot}yr^{-1}$ \\
\hline && \\[-4pt]
Globular cluster & $\le10^8$ & $\ge 0.01$ \\
$\omega$Cen&$\ge10^8$ & $\le 0.1$ \\
Halo, [Fe/H]$\le-2.0$ & $\le10^8$ & $\sim 1$ \\
Halo, [Fe/H]$\sim-1.5$ & $\le10^9$ &  $\sim 1$ \\
Bulge; high [$\alpha$/Fe] & few.10$^8$ & 10-100 \\
Bulge; low [$\alpha$/Fe] & few.10$^9$ & 10-100 \\
Thick Disk & few.10$^9$ & 1-10 \\
Current Disk & $10^{10}$ & $\sim 1-10$ \\
Inner Disk & ? & ? \\
Satellite dSph & many.10$^9$ & $\le 10^{-3}$ \\
\hline && \\[-4pt]
 Assembly & early & \\
\hline && \\[-4pt]
Infall & continuing? & $\sim 4$Gyr \\
\hline
\end{tabular}
\vspace{-0.5cm}
\end{center}
\end{table}

In cases where no spread in element ratios is seen, and there is no
range in [Fe/H], star formation was plausibly complete before new
chemical elements could be produced; perhaps globular clsuters are an
example of this case. For field halo stars with [Fe/H]$\ge -2$, where
a wide range in [Fe/H] but a very small range in [$\alpha$/H] is seen,
star formation must have continued for long enough for efficient
mixing of supernova ejecta into the ISM. Since no products of Type~I
supernovae are seen, this brackets allowed star formation
durations. By applying such qualitative considerations, we deduce that
most of the Milky way formed at a star formation rate which is
comparable to that of today. Only the Galactic Bulge, and the inner
disk, where star formation histories remain very poorly known, are
available to retrieve the Milky Way's place as a `typical object' on
the Madau plot. 
These simple calculations are summarised in Table~(1).

\section{ Quantitative Determination of Star Formation Rates}

Most attempts at deducing the past history of the star formation rate
in the Galaxy have relied on indirect age indicators, typically
chemical evolution models, using an age-metallicity relation, or from
empirical correlations of stellar properties, such as chromospheric
activity, with age. Because of the many uncertainties and
observational biases inherent in these methods, many corrections must
be applied to the samples before a SFH can be inferred. A good recent
example, illustrating the strengths and complexities of the
techniques, is Rocha-Pinto et al. (2000). Given all this, it is not
too surprising to find conflicting results from independent analyses
of the limited available data.

The precision in luminosities of field stars provided by the Hipparcos
satellite gives us for the first time the data quality to allow
application of a more objective method to infer the star formation
history which produced the solar neighbourhood. Indeed, recent
analytical developments show that it is now possible to reconstruct
the star formation history which gave rise to the observed
distribution of stars in the HR diagram for any (relatively) simple
stellar population. The constraints in present application of the
methodology are that one requires (i) a small scatter in the
metallicity distribution, and known mean abundance, (ii) a
colour-magnitude diagram extending below the turnoff age of interest,
and (iii) appropriate stellar evolutionary tracks. Whilst previous
reconstructions of the star formation history based on the inversion
of CMDs were forced to rely on assumed parametric forms for the
evolution of the star formation rate, we have developed and
implemented a new and rigorous method which makes no {\sl a priori}
assumptions about the possible complexity of the star formation history.

This method is explained in detail in Hernandez et al. (1999, Paper~I)
and in Gilmore etal. (2000). These references also present results
from the extensive simulations which were used to quantify the biases
inherent in any reconstruction based on observational colour-magnitude
data.  The effects of unresolved binaries, of uncertainties in the
stellar initial mass function, and of errors in the adopted system
metallicity, were assessed. It was concluded that derivation of an
absolute normalisation of the derived star formation rate is still not
robust, given the uncertainties in the distribution of mass ratios in
binaries, of the fraction of binaries in a given sample, of the IMF,
and of metallicity distributions. However, a relative star formation
history can be inferred, where the relative amplitudes of the
variations in star formation rate are correct, provided the IMF, the
properties of binaries and the distribution of metallicity do not
evolve with time. With these provisos, the method was applied to a
sample of HST CMDs of dSph galaxies of the Local Group (Hernandez,
Gilmore, \& Valls-Gabaud 2000, Paper~II). A further paper extends the
application to the Solar neighbourhood (Hernandez, Valls-Gabaud \&
Gilmore 2000, Paper~III).

Full details of the method are presented in Paper~I, and in Gilmore et
al. (2000), and will not be repeated here.  Very briefly, the method
takes as inputs only the positions of $n$ stars in a colour magnitude
array, each having a colour $c_i$ and luminosity $l_i$, with
(uncorrelated) associated errors $\sigma(c_i)$ and $\sigma(l_i)$
respectively.  Using the likelihood technique, we first construct the
probability that the $n$ observed stars resulted from some function
$SFR(t)$. This is given by

\begin{equation}
{\cal L}= \prod_{i=1}^{n} \left( 
\int_{t_0} ^{t_1} \, SFR(t) \, G_{i}(t) \, dt \right),
\end{equation}

\noindent where

\begin{eqnarray}
G_{i}(t)  =  \int_{m_0}^{m_1} {\rho(m;t) \over{2 \pi \sigma(l_i)
 \sigma(c_i)}} \,  
 \exp\left(-D(l_{i};t,m)^2 \over {2 \sigma^2(l_i)} \right) 
\exp\left(-D(c_{i};t,m)^2 \over {2 \sigma^2(c_i)} \right) \hspace{5pt} 
 dm 
\end{eqnarray}

In the above expression $\rho(m;t)$ is the density of points along the
isochrone of age $t$, around the star of mass $m$. It is determined by
convolving an assumed IMF with a stellar evolutionary track, defining
the duration of the differential evolutionary phases for a star of
mass $m$. The ages $t_0$ and $t_1$ are a minimum and a maximum age
which need to be considered in the specific problem of interest, while
$m_0$ and $m_1$ are a minimum and a maximum mass which need to be
considered along each isochrone, typically 0.6${\mathcal M}_{\odot}$
and 20${\mathcal M}_{\odot}$.  $D(l_{i};t,m)$ and $D(c_{i};t,m)$ are
the differences in luminosity and colour, respectively, between the
$i$th observed star and a general star of age and mass $(m,t)$.  We
refer to $G_{i}(t)$ as the likelihood matrix, since each element
represents the probability that a given star, $i$, was actually formed
at time $t$ with any mass.

Following the discussion of Paper~I, we may write the condition that
the likelihood has an extremal as the variation $\delta {\cal L}(SFR)
= 0 $, allowing a full variational calculus analysis to be
applied. Developing first the product over $i$ using the chain rule,
and dividing the resulting sum by ${\cal L}$, one obtains:

\begin{equation}
\sum_{i=1}^{n} \left(
{\delta \int_{t_0} ^{t_1} SFR(t) \, G_{i}(t) \, dt} \over {\int_{t_0}^{t_1}
 SFR(t) \, G_{i}(t) \, dt} \right) =0
\end{equation}

\noindent Introducing the new variable $Y(t)$ defined as:

$$
Y(t)=\int{ \sqrt {SFR(t)} \, dt}\;  \Longrightarrow \;  SFR(t)=\left( {dY(t)
\over dt} \right)^2
$$

\noindent
into Equation~(2) we can develop the Euler equation to yield 

\begin{equation}
{d^2 Y(t)\over dt^2}\sum_{i=1}^{n} \left( G_{i}(t) \over I(i)\right)
=-{dY(t)\over dt}\sum_{i=1}^{n} \left( dG_{i}/dt \over I(i)\right)
\end{equation}

\noindent where 
$$
I(i)=\int_{t_0}^{t_1} SFR(t)\,  G_{i}(t) \, dt
$$
\noindent
We have now transformed what was an optimisation problem, finding the
function that maximises the product of integrals defined by
equation~(1), into an integro-differential equation with a boundary
condition (at either $t_o$ or $t_1$) which can be solved by iteration.
Further details on the numerical aspects of the procedure are
available in Paper~I, and need not be repeated here.

This methodology has two important advantages over traditional maximum
likelihood techniques: \\ (i) variational calculus allows a fully
non-parametric approach, independent of one's astrophysical
pre-conceptions; and \\ (ii) since the optimal star formation history
$SFR(t)$ is solved for directly, the computational procedure is very
fast, not requiring repeated CPU intensive comparisons between
observed and synthetic diagrams.

\section{Objective Determination of Recent Star Formation Histories}

We now apply the methodolgy outlined above to the Solar
Neighbourhood. A suitable data set can be derived from the Hipparcos
catalogue (ESA 1997), providing absolute luminosities and colours
which are, with careful selection, are almost bias-free. The sample is
however restricted to bright stars, and so samples only a very small
volume, and releatively young ages. We have derived from the Hipparcos
catalogue a volume-limited survey of the solar neighbourhood. This
selection is described in detail in Paper~III, along with the
kinematic and geometrical corrections that have to be made to correct
the sample for its selection function.

\begin{figure}[h]
\centerline{\psfig{file=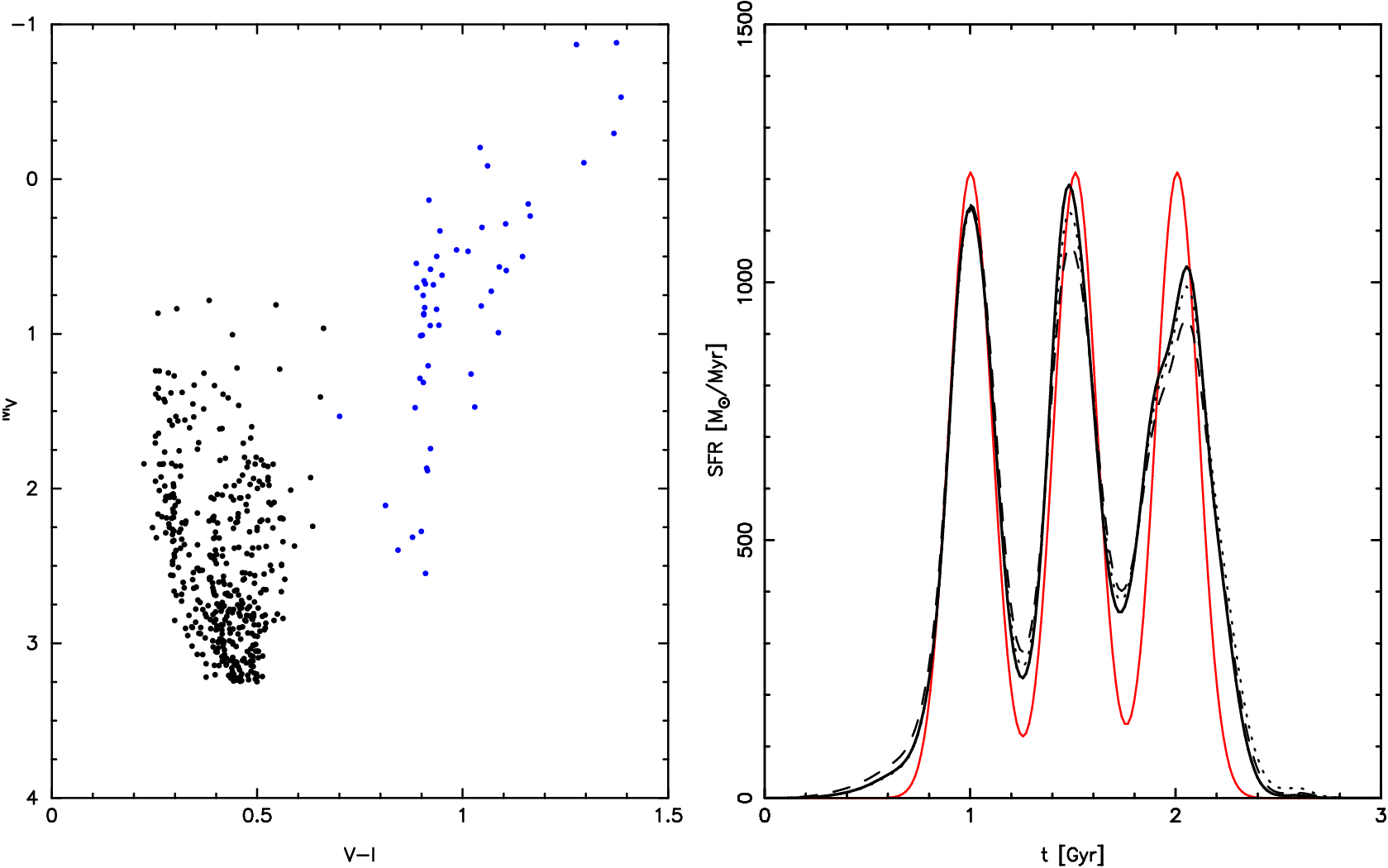,height=6cm,width=12cm,angle=-90.}}
\caption[]{Proving the method with a series of 3 bursts of star
formation. The synthetic CMD (left panel) results from the 3-burst
star formation history shown by the dotted line in the right panel.
The solid lines on the right panel are the last three iterations of
the inversion, given no information except the CMD of the left panel,
and an appropriate metallicity.  An accurate reconstruction of the
star formation history is apparent.  This simulation utilises the same
number of stars and the same observational uncertainties as in the
Hipparcos sample. }

\psfig{file=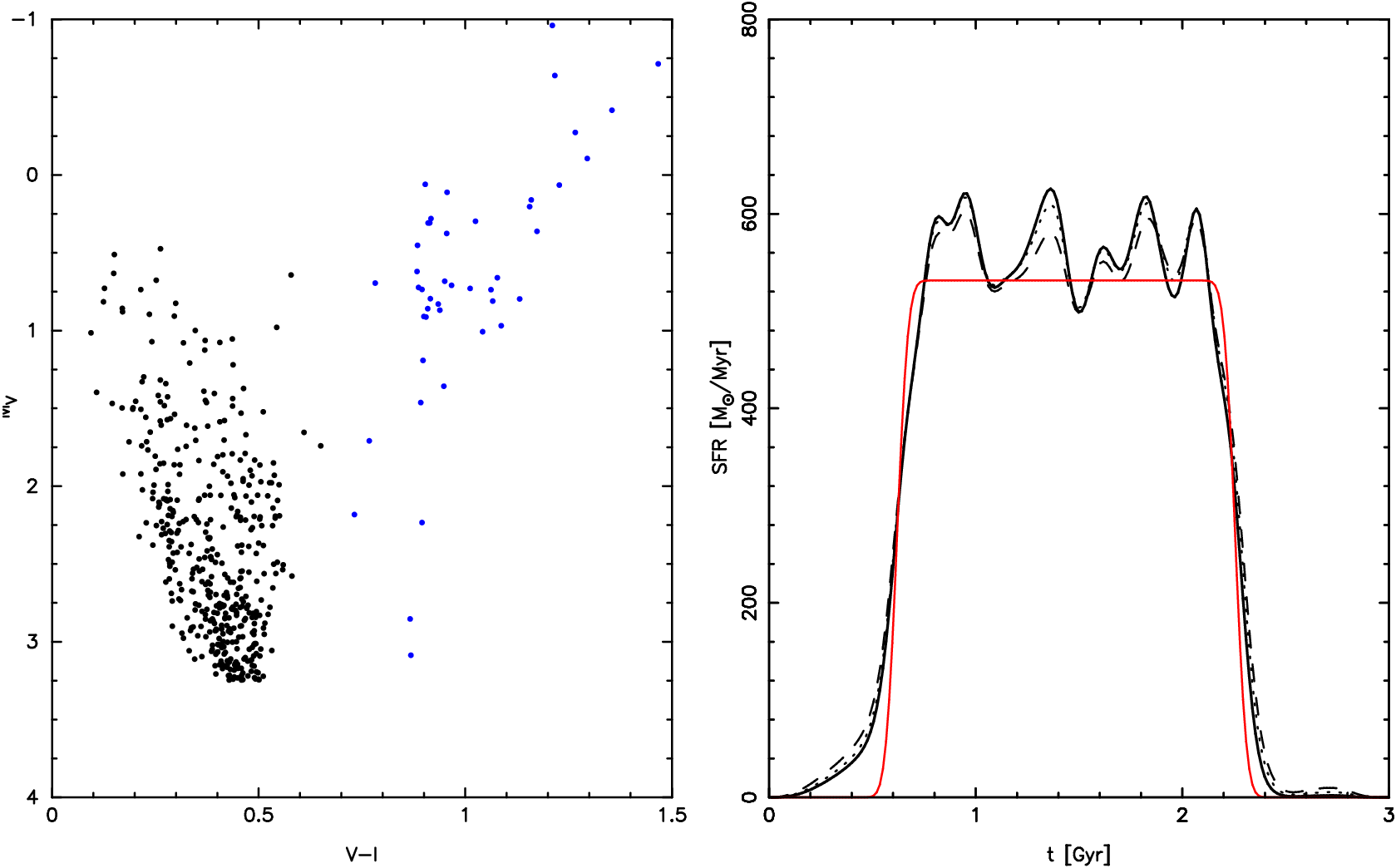,height=6cm,width=12cm,angle=-90.}
\caption[]{Same as Figure~1, except that the simulated CMD is derived
from a constant SFR over several Gyr (dotted curve, right hand
panel). Note that the CMD appears superficially very similar to the
one presented in Fig.~1, even though the simulated star formation
history is quite different.  The method implemented here is able
to distinguish and recover the correct (test) star formation
histories.}
\end{figure}

In summary, we restrict consideration to non-variable stars with
apparent magnitude brighter than $V=7.9$ and error in parallax smaller
than 20\%. To avoid a wide distribution in the uncertainties in colour
and luminosity the sample is further reduced to absolute magnitudes
brighter than $M_V =3.15$.  This absolute magnitude-limited sample
implies that only stars younger than about 3 Gyr are
considered. Determination of the star formation history of the
Galactic disk at earlier times awaits a deeper sample, such as will be
naturally provided by the GAIA mission.

\subsection{Simulations and calibrations}

This analysis extends the temporal resolution of the method to much
shorter times than used in the simulations described in Paper~I.  In
order to check whether samples such as the one selected can in fact be
inverted to infer their star formation history with extremely high
temporal resolution, we performed a further extensive set of of
tests. As before, these involved creation, convolution with an error
and sampling function, and objective inversion, of colour magnitude
data resulting from a variety of complex star formation
histories. Figure~1 illustrates one such test, where a series of 3
bursts (right panel, dashed line) gives rise to the synthetic CMD
shown on the left.  The solid lines in the right panel give the last 3
iterations of the method, showing convergence to an accurate
reconstruction of the input star formation history.

Similarly, Figure~2 shows a test with a constant star formation rate
continuing over several Gyrs. The small fluctuations in the
reconstructed star formation history arise from numerical
instabilities created by the shot noise due to the small number of
stars involved, about 450. This implies that a smoothing procedure has
to be applied, resulting in a degradation in the effective time
resolution to about 50 Myr. Note also that only stars bluewards of
$V-I$=0.7 are used, to avoid unnecessary complications created by a
small number of potentially older stars with an extremely poorly
quantified selection function contaminating the reconstruction.

\section{Star Formation History of the Solar Neighbourhood}

Figure~(3) shows the colour magnitude diagram for those stars in the
volume-limited sample complete to $M_{V}<3.15$ for stars in the
Hipparcos catalogue having errors in parallax of less than $20 \%$ and
with apparent magnitude $m_{V}<7.25$ (left panel). The right panel of
this figure shows the result of applying our inversion method to this
data set. The dotted envelope spanning the best-fit star formation
history quantifies the range of many reconstructions arising from
different $M_{V}$ selection cuts in the Hipparcos catalogue.  This
quantifies the plausible range of systematic errors likely to be
present, due predominately to the small sample of stars, and the
inevitably {\sl a posteriori} selection function which must be applied
to Hipparcos data.  Reconstruction based on the $(M_V, B-V)$ diagram
rather than $(M_V,V-I)$ gives essentially the same results, showing
that the available isochrones are a good match to the photometric
systems.

\begin{figure}[h]
\psfig{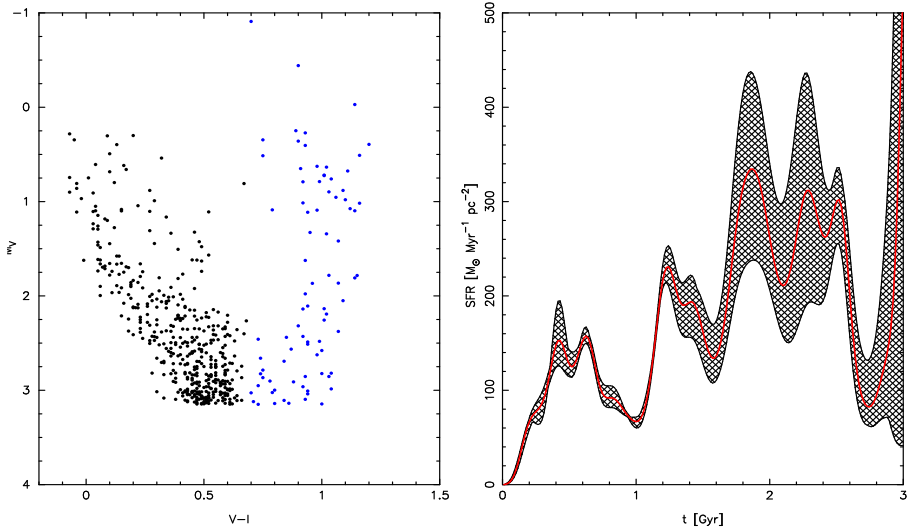}
\caption[]{The left panel shows the colour magnitude data for the
complete sample of nearby massive stars with good parallaxes from
Hipparcos.  The right panel presents a best-fit reconstruction of the
star formation history (solid line), along with the envelope resulting
from reconstructions with different $M_V$ cuts, to assess the
robustness of the inferred star formation history. The decrease to
zero at recent times is an artefact of the very small sampling volume.}
\end{figure}   

The derived local star formation history of the Solar Neighbourhood
shown in Figure~(3) shows what may be described as an underlying
constant level of star formation activity, onto which is superimposed
a strong, quasi-periodic variation with a period close to $0.5$ Gyr.
The high time resolution of our star formation history
reconstruction makes it difficult to compare with the results derived
from chromospheric activity studies (eg Rocha-Pinto et al. 1999),
although qualitatively we do find the same activity at both 0.5 and
above 2 Gyr, but not the decrease found by them between 1 and 2 Gyr.

Assuming the variable component really is a cyclic pattern
superimposed on a base level, and that our brief time interval is
sufficient to identify a true characteristic `period', we may consider
its meaning.  One possible interpretation of a cyclic component, or at
least of some temporal regularity in the star formation history of the
solar neighbourhood, can be found in the density wave hypothesis (Lin
and Shu, 1964) for the presence of spiral arms in late type
galaxies. As the pattern speed and the circular velocity are in
general different, the solar neighbourhood periodically crosses an arm
region, where the increased local gravitational potential might
possibly trigger an episode of star formation. In this case, the time
interval $\Delta t$ between encounters with an arm at the solar
neighbourhood is

$$
\Delta t \; = \; \frac{0.22 \, {\rm Gyr}}{m} \; \left(\frac{\Omega}{29 \,
 {\rm km \, s}^{-1} {\rm kpc}^{-1}}\right)^{-1} \; 
 \left| \frac{\Omega_p}{\Omega} - 1  \right|^{-1}
$$
 
\noindent
where $m$ is the number of arms in the spiral pattern. The classical
value of the pattern speed, $\Omega_p = 0.5 \, \Omega \approx 14.5$km
s$^{-1}$ kpc$^{-1}$ would imply the rather surprising conclusion that 
interaction with a single arm ($m=1$) would be enough to account
for the observed regularity in the recent SFR history. 

However, more recent determinations tend to point to much larger
values for the pattern speed (e.g.  Mishkurov et al. 1979, Avedisova
1989, Amaral and L\'epine 1997) close to $\Omega_p \sim 23 - 24$ km
s$^{-1}$ kpc$^{-1}$, which would then imply that the regularity
present in the reconstructed $SFR(t)$ would be consistent with a
scenario where the interaction of the solar neighbourhood with a
two-armed spiral pattern would have induced the star formation
episodes we detect. This is reminiscent of the explanations put
forward to account for the inhomogeneities observed in the HIPPARCOS
velocity distribution function, where well-defined branches associated
with moving groups of different ages (Chereul et al. 1999, Skuljan et
al. 1999, Asiain et al. 1999) could perhaps be also associated with an
interaction with spiral arm(s), although in this case the time scales
are much smaller.  Of course, other explanations are possible; for
example the cloud formation, collision and stellar feedback models of
Vazquez \& Scalo (1989) predict a phase of oscillatory star formation
rate behaviour as a result of a self-regulated star formation
r\'egime. Close encounters with the Magellanic Clouds have also been
suggested to explain the intermittent nature of the star formation
rate, though on longer time scales (Rocha-Pinto et al. 2000).

\subsection{Are the Results Reliable?}

The first question that arises upon application of any new technique,
especially one involving complex numerical optimisations, is how
reliable is the reconstruction of the star formation history we have
deduced?  The most common procedure to compare a specific star
formation history with an observed CMD has been to use the star
formation history to generate a synthetic CMD, and then to compare
this to the observations, using some statistical test to determine the
degree of similarity between the two.

This method has a significant disadvantage in the present case, in
that one is not comparing the `true' star formation history with the
data, but rather a particular realisation of the star formation
history is being compared with the data.  The distinction becomes
arbitrary only when large numbers of stars are found in all regions of the
colour magnitude diagram. This is not true in general, and especially
not so in this application. Instead we follow a Bayesian approach,
and adopt for use here the $W$ statistic presented by Saha (1998).
This is essentially

$$
W=\prod_{i=1}^{B} {{(m_{i}+s_{i})!} \over {m_{i}!s_{i}!}}
$$

\noindent
where B is the number of cells into which the CMD is split, and
$m_{i}$ and $s_{i}$ are occupation numbers, the numbers of points
which the two distributions being compared have in each cell. This
asks for the probability that two distinct data sets are random
realisations of the same underling distribution.

\begin{figure}[hb]
\psfig{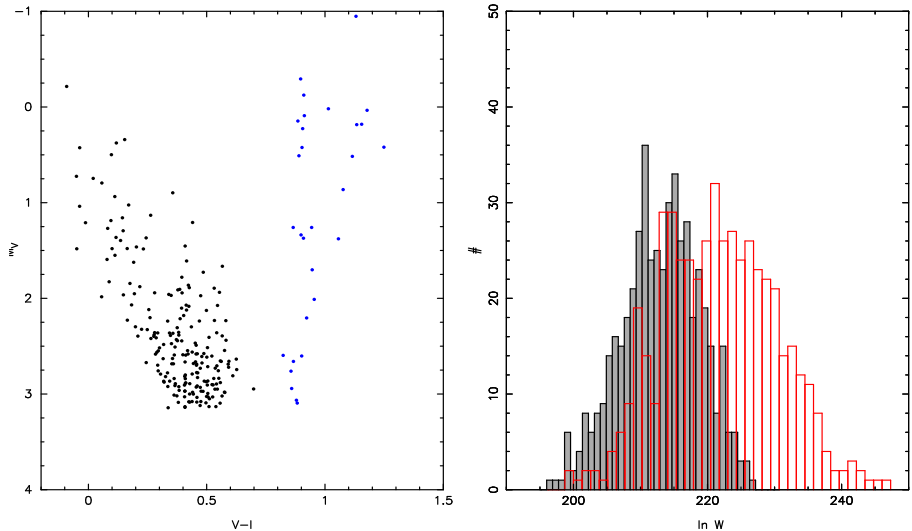}
\caption[]{The left panel shows one synthetic colour magnitude
distribution resulting from the reconstructed star formation history
of Fig.~3 (left panel), generated with the same selection function as
for the observed Hipparcos sample. The right panel shows the
distribution of the $W$ statistic for 500 model-model comparisons
(solid lines), illustrating the range of the $W$ statistic expected
from sampling statistics, given a fixed underlying star formation
function.  The dashed line shows the $W$ statistics resulting from 500
comparisons of simulated models and the actual Hipparcos data, from
Figure~(3). The overlap between the two sets of results shows that the
derived star formation history of Figure~(3) is indeed a statistically
valid representation of the data. }
\end{figure}

In implementing this test we first produce a large number of random
realisations of the colour magnitude diagram generated by our inferred
star formation history, and then compute the $W$ statistic between
pairs in this sample of CMDs. This gives a distribution which is used
to determine the range of values of $W$ which is expected to arise in
random realisations of the star formation history being tested. Next,
we produce a new set of a large number of random realisations of the
colour magnitude diagram generated by our inferred star formation
history, and compute the $W$ statistics between the observed data set
and these new random realisations of the star formation history. This
gives a new distribution of $W$, which can then be objectively
compared to the one arising from the model-model comparison. This
allows us to assess whether both data and modeled colour magnitude
diagrams are compatible with a unique underling distribution.

Figure~(4) shows one such synthetic colour magnitude diagram produced
from our inferred star formation history for the solar neighbourhood,
down to $M_{V}=3.15$. This can be compared to the Hipparcos CMD
complete to the same $M_{V}$ limit shown in Figure~(3). A visual
inspection reveals approximately equal numbers of stars in each of the
distinct regions of the diagram. Such a comparison is of little value,
however, as noted in dicussion of the simulations summarised in
Figures~(1) \& (2).  A rigorous statistical comparison is also
possible. The right panel of Figure~(4) shows a histogram of the
values of the $W$ statistic for 500 random realisations of our
inferred star formation history in a model-model comparison. This
gives the probability density function for the $W$, given our star
formation history. The dashed histogram presents the result of the
comparison of 500 synthetic colour magnitude diagrams with the
observed Hipparcos data set: both sets of $W$ are compatible.

\section{Conclusion}

Star formation rates and histories can be estimated in special cases
from a combination of chemical evolution models and the total stellar
mass formed into stars. Basically, this exploits the stellar
evolutionary, and Type~I supernova timescales for element production.
In cases where no spread in element ratios is seen, and there is no
range in [Fe/H], star formation was plausibly complete before new
chemical elements could be produced; perhaps globular clsuters are an
example of this case. For field halo stars with [Fe/H]$\ge -2$, where
a wide range in [Fe/H] but a very small range in [$\alpha$/H] is seen,
star formation must have continued for long enough for efficient
mixing of supernova ejects into the ISM. Since no products of Type~I
supernovae are seen, this brackets allowed star formation
durations. By applying such qualitative considerations, we deduce that
most of the Milky way formed at a star formation rate which is
comparable to that of today. Only the Galactic Bulge, and the inner
disk, where star formation histories remain very poorly known, are
available to retrieve the Milky Way's place as a `typical object' on
the Madau plot.

In the immediate Solar neighbourhood, and in dSph satellite galaxies,
the high quality data and the simple stellar populations respectively
allow us to be more quantitative.  We have applied the objective
variational calculus method for the reconstruction of star formation
histories from observed colour magnitude data, developed in our
Paper~(I), to the data in the Hipparcos catalogue, yielding the star
formation history of the solar neighbourhood over the last 3
Gyr. Surprisingly, a structured star formation history is obtained,
showing a cyclic pattern with a period of about 0.5 Gyr, superimposed
on some underlying star formation activity which increases slightly
with age.  No random bursting behaviour was found at the time
resolution of 0.05 Gyr of our method. A first order density wave model
for the repeated encounter of galactic arms could explain the observed
regularity.

\end{document}